
\def\kr{cosmic ray }
\def\krs{cosmic rays }
\def\nuci{nuclei }
\def\nuc{nucleus }

\MAINTITLE={Interactions of \kr \nuci}
\AUTHOR={K. Mannheim$^1$ and R. Schlickeiser$^2$}
\INSTITUTE={$^1$ Universit\"ats-Sternwarte, Geismarlandstr. 11,
D-37083 G\"ottingen, Germany\newline $^2$
Max--Planck--Institut f\"ur Radioastronomie,
Auf dem H\"ugel 69,
D--53121 Bonn,
Germany}
\ABSTRACT={Convenient formulae for the energy losses of energetic atomic nuclei
over the entire energy range relevant to the physics
of cosmic rays are presented.  Results can readily
be applied to transport equations of cosmic rays under
quite general conditions as demonstrated by solving
a leaky--box equation with a  complete loss term.
Thereby we derive the equilibrium spectrum of cosmic rays
in various types of galaxies.
We emphasize a spectral break energy at
$T^*\approx 450$~MeV independent of the matter density, resulting from the
transition
from Coulomb and ionization losses to pion production losses as
the relevant cooling process for the cosmic ray \nuci .  We comment on the
possible
cosmic ray origin of the cosmic gamma ray background.}
\KEYWORDS={Cosmic rays -- Gamma rays: theory -- Interstellar medium: general
-- Nuclear reactions -- Galaxies: active}
\THESAURUS={02.14.1 -- 09.03.2 -- 09.07.1 -- 11.01.2 -- 13.07.3}
\OFFPRINTS={K. Mannheim}
\DATE={}
\maketitle
\titlea{Introduction}
Most probably the gamma ray properties of
galaxies are indicative for the presence of baryonic
cosmic rays.  Therefore,
recent advances in gamma ray astronomy demand for a
compilation of energy loss calculations for cosmic ray
nuclei.  Hence one can determine the stationary cosmic
ray spectrum and the resulting gamma ray or neutrino emissivity.

Pointlike electromagnetic interactions of cosmic ray electrons
(synchrotron radiation, inverse Compton scattering,
nonthermal brems\-strahlung)
involve the Thomson cross section being proportional to the square
of the electron radius
of the radiating particle. Therefore,
if we consider a heavy \kr nucleus of
charge Ze and mass $M=Am_{\rm p}$ instead of a \kr electron,
the corresponding cross sections involve the square of the nucleon
radius $R_\circ=(Ze)^2/(Am_{\rm p}c^2)=(Z^2/A)(m_{\rm e}/m_{\rm p})r_\circ$
yielding the
Thomson cross section for \nuci
$$\sigma _N=(Z^4/A^2)(m_{\rm e}/m_{\rm p})^2\sigma _T=1.97\cdot 10^{-31}
{{Z^4}\over {A^2}}\, {\rm cm}^2\ \eqno
(1.1).$$
As a consequence
cross sections of pointlike
electromagnetic interactions of \kr \nuci are much
smaller than those of \kr electrons of the same Lorentz factor,
and can be neglected
in almost all astrophysical applications as production processes
for cosmic photons and as energy loss processes for {\kr \nuci}. A noteworthy
exception is the limit on the  attainable energy for \kr \nuci in pulsar
acceleration models set by heavy curvature radiation losses (Sorrell 1987).

The remaining inelastic
interaction processes can be divided into two groups:

\item{(i)} interactions with photons,

\item{(ii)} interactions with matter.

\noindent
Before we consider each in turn, we
summarize the relevant kinematics of inelastic
collisions.
Finally, we
conclude by solving
a transport equation with energy--dependent escape--time
and a complete loss--term for hydrogen nuclei.
Thereby we obtain a description of
the stationary spectrum of \kr  protons in various types of
galactic environments.
\titlea{Relativistic kinematics of inelastic collisions}
\titleb{Threshold energy}
Consider the collision of two particles a and b leading to
the creation of particles c,d,... in the laboratory system
$$\rm a+b\to c+d+.....\eqno (2.1),$$
where the particles have masses
$m_{\rm a}$
and $m_{\rm b}$ and
four--momenta
$P_{\rm a}=({\varepsilon _{\rm a}\over c},\vec p_{\rm a})$ and
$P_{\rm b}=({\varepsilon _{\rm b}\over c},\vec p_{\rm b})$, respectively.
To create new particles in this reaction the minimum energy of the incoming
particles a and b has to be equivalent to just enough available energy in
the center--of--momentum system (CMS\fonote{CMS
quantities are denoted with a prime.}) to produce the rest mass of all
outgoing particles c,d,....
{}From the invariance of the four--momentum
we obtain for the total energy $E^\prime$ in the CMS
$$E^{\prime 2}
=c^2(P_{\rm a}+P_{\rm b})^2=
m_{\rm a}^2c^4+m_{\rm b}^2c^4
+2\varepsilon _{\rm a}\varepsilon _{\rm b}-2\vec p_{\rm a} \vec p _{\rm b}c^2
\eqno(2.2).$$
The threshold energy for the creation of particles c,d,... in reaction (2.1)
is
$$E_{\rm th}^\prime =m_{\rm a}c^2+m_{\rm b}c^2+\Delta mc^2
=m_{\rm c}c^2+m_{\rm d}c^2+....\eqno (2.3)$$
where $\Delta m$ is the mass difference between incoming and outgoing
particles. Combining Eq.(2.2) and (2.3) we obtain
the relation
$$\eqalign{&\varepsilon _{\rm a} \varepsilon _{\rm b}-\vec p_{\rm a}\vec p_{\rm
b}c^2=\cr &
m_{\rm a}m_{\rm b}c^4+\Delta mc^4\left(m_{\rm a}+m_{\rm b}+{\Delta m\over
2}\right)\cr}\eqno (2.4).$$
In case both incoming particles have non--zero masses Eq.(2.4) reduces to
$$\eqalign{&\gamma _{\rm a}\gamma _{\rm b}-\sqrt{(\gamma _{\rm a}^2-1)(\gamma
_{\rm b}^2-1)}\cos \theta
\cr &=1+\Delta m\left({1\over m_{\rm a}}+{1\over m_{\rm b}}+{\Delta m\over
{2m_{\rm a}m_{\rm b}}}\right)
\cr}\eqno (2.5),$$
where we introduced the lab--system Lorentz factors of the particles
and where $\theta $ denotes the collision angle of the reaction in
the lab--system.

In case particle b is a photon ($m_{\rm b}$=0), Eq.(2.5) reduces to
$$\varepsilon _{\rm b}\left(\gamma _{\rm a}-\sqrt{\gamma _{\rm a}^2-1}\cos
\theta\right)
=\Delta mc^2\left(1+{\Delta m\over {2m_{\rm a}}}\right)\eqno (2.6).$$
\titleb{The energy of one particle seen from the rest system of another
one}
Another useful calculation concerns the energy of particle b in reaction
(2.1) seen from the rest system of particle a. We call that energy
$E_{ba}$. Using the invariance of $P_{\rm a}P_{\rm b}$ we find
$$P_{\rm a}P_{\rm b}={\varepsilon _{\rm a}\varepsilon _{\rm b}\over {c^2}}-\vec
p_{\rm a}\vec p_{\rm b}=m_{\rm a}E_{ba}
\eqno (2.7).$$
In case both incoming particles have non--zero masses Eq.(2.7)
reduces to
$$E_{ba}=m_{\rm b}c^2\left[\gamma _{\rm a}\gamma _{\rm b}-
\sqrt {(\gamma _{\rm a}^2-1)(\gamma _{\rm b}^2-1)}\cos \theta \right]\eqno
(2.8),$$
whereas in case particle b is a photon we obtain
$$E_{ba}=\varepsilon _{\rm b}\left[\gamma _{\rm a} -\sqrt {\gamma _{\rm
a}^2-1}\cos \theta\right]
\eqno (2.9).$$
\titlea{Interactions between \kr \nuci and  photons }
Three basic interactions have important astrophysical consequences for
relativistic \nuci : (1) pair production (particularly of $e^+e^- $ pairs)
in the field of the nucleus $\rm A+ \gamma \to A+e^++e^-$, (2) photoproduction
of hadrons (mostly pions) $\rm A+\gamma \to A+\pi 's$ , and (3)
photodisintegration of the nucleus $ \rm A+\gamma \to (A-1)+N$
where $\rm N=p,n$.
As will be shown further in the text, when process (2)
becomes important only protons need to be considered,
since due to process (3) nuclei do not survive under such conditions.
In the rest--system
of the \kr \nuc we find according to Eq.(2.4) with $\vec p_{\rm a}=0$,
$\varepsilon _{\rm a}=m_{\rm a}c^2$, and $m_{\rm b}=0$ for the photon threshold
energy
$$\varepsilon _{\rm th}^\prime =\Delta mc^2\left[1+{\Delta m\over {2m_{\rm
a}}}\right]\eqno (2.10).$$
Process (1) ($\Delta m=2m_{\rm e}$) then occurs for photon energies above
$$\varepsilon _{th,e^+e^-}^\prime =2m_{\rm e}c^2\left(1+{m_{\rm e}\over
{Am_{\rm p}}}\right)\simeq
2m_{\rm e}c^2\simeq 1\thinspace {\rm MeV}\eqno (2.11),$$
and process (2) at
$$\varepsilon _{\rm th,\xi\pi }^{\prime
}={\rm \xi}m_{\pi }c^2\left[1+{{{\rm \xi}m_{\pi }}\over {2m_{\rm p}}}\right]\
\eqno (2.12),$$
for the production of $\xi$ pions so that for the production of a single pion
the rest--system threshold energy is $\varepsilon _{th,\pi }^\prime =
145$~MeV.
Process (3) is particularly important in most cases between 15
and 25~MeV where the giant dipole resonance has its peak (Puget {\it {\it et
al.}}
1976).

If the target photons are distributed isotropically in space and have mean
energy $\langle\varepsilon \rangle$ in the observer's frame of reference the
minimum
threshold Lorentz factor of the relativistic ($\gamma _{\rm a}\gg 1$) \kr \nuc
according to Eq.(2.6) occurs for a head--on collision ($\cos \theta =-1$)
and is
$$\gamma _{min}={\Delta mc^2\over {2\langle\varepsilon \rangle}}\left[1+{\Delta
m\over {2m_{\rm a}}}\right]
\eqno (2.13).$$
In intergalactic space
the most prominent radiation field is the microwave
background radiation with $\langle\varepsilon \rangle=7\cdot 10^{-4}$~eV, so
that processes
(1) and (2) occur only for ultrahigh energy {\kr \nuci} (Greisen 1966),
$$\gamma _{min}^{(1)}
\simeq m_{\rm e}c^2/\langle\varepsilon \rangle=7\cdot 10^8\eqno (2.14),$$
and
$$\gamma _{min}^{(2)}\simeq
{{\rm \xi}m_{\pi }c^2\over {2\langle\varepsilon \rangle}}\left(1+{{\rm
\xi}m_{\pi }\over {2m_{\rm p}}}\right)\simeq
{\rm \xi}\cdot 10^{11}\eqno (2.15),$$
respectively.
Pair production and photoproduction interactions with
extragalactic optical and infrared photons are negligibly small, because their
radiation density in the intergalactic medium is so small, that the respective
time scale
of cooling is longer than the Hubble time scale.
\titleb{Photo--pair production}
Bethe-Heitler pair production (Bethe \& Heitler 1934)
in the field of the nucleus
in an isotropic radiation field has been treated in detail
in its astrophysical context by Blumenthal (1970),Berezinsky \&
Grigoreva (1988) and Chodorowski {\it et al.} (1992).
The energy loss rate for a \kr \nuc
of charge $Ze$ and Lorentz factor $\gamma$ is
$$\eqalign{&-\left({dE\over {dt}}\right)_{\rm e^\pm}={3c\alpha_{\rm f} \sigma
_T\over {8\pi }}Z^2(m_{\rm e}c^2)^2
\cr &\cdot
\int_2^{\infty }dx n_\gamma\left({x m_{\rm e}c^2\over {2\gamma }}\right)
{\phi (x )\over {x ^2}}\cr}
\eqno (3.1),$$
where $m_{\rm e}$ is the electron mass, $\alpha_{\rm f}=1/137$ is the fine
structure constant,
$\sigma _T=6.65\cdot 10^{-25}{\rm cm}^2$ is the electron Thomson cross
section,
$\phi (x )$ is an integral over the pair--production cross
section and $n_\gamma(\varepsilon )$ is the number density distribution of the
target photons. The asymptotic formula for $x \gg  1$,
$$\eqalign{&\phi (x \gg  1)\to x [-86.07 +\
50.95\ln x \ -14.45(\ln x )^2\cr &
+\ 2.667(\ln x )^3]\cr}\eqno (3.2)$$
works reasonably well for all values of $x \ge 2$ in Eq.(3.1) when the
target photon distribution $n_\gamma(\varepsilon )$ is such that for all
contributing
photons $\gamma \varepsilon \gg  m_{\rm e}c^2$. With the graybody distribution
function
we obtain for \kr protons
$$-\left({dE\over {dt}}\right)_{\rm e^\pm}={45\ c\ u_\gamma\ \alpha_{\rm f}
\sigma _T(m_{\rm e}c^2)^2
\over {8\pi ^5\ (kT)^2}}f\left({m_{\rm e}c^2\over {2\gamma kT}}\right)\eqno
(3.3),$$
where
$$f(\nu )\equiv \nu ^2\int _2^{\infty }dx \phi (x )(e^{\nu x }
-\ 1)^{-1}\eqno (3.4).$$
$f(\nu )$ has been
calculated by Blumenthal (1970) and from his figure of the result
it is clear that
the threshold for
pair production is at $\nu \approx 1$ corresponding to
$\gamma \approx {m_{\rm e}c^2\over {2kT}}$ in agreement with (2.14), since
for a graybody distribution $\langle\varepsilon \rangle=2.7$~kT.
{}For the special case of the microwave background blackbody radiation
field one may express the photon energy density in terms of the blackbody
temperature according to the Stefan-Boltzmann law,
$u_\gamma=a T^4$ where $a =7.56\cdot 10^{-15}$~erg~cm$^{-3}$~K$^{-4}$.
As an example we obtain the energy loss rate for protons at threshold
$$\left(dE_{\rm p}\over dt\right)_{\rm e^\pm}=2.4\cdot 10^5\left(u_\gamma\over
{\rm erg\thinspace cm^{-3}}\right)
\left(\langle kT\rangle\over {\rm eV}\right)^{-2}\ {\rm eV\over s}\eqno(3.5).$$
In a power law photon distribution with $n_\gamma(\varepsilon)\propto
\varepsilon^{-2}$ the
energy loss rate for all proton energies $E_{\rm p}$
becomes
$$\left(dE_{\rm p}\over dt\right)_{\rm e^\pm}=3.1\cdot
10^{-7}\left(u_\gamma\over {\rm erg\thinspace cm^{-3}}\right)
\left(E_{\rm p}\over {\rm GeV}\right)^2\ {\rm eV\over s}\eqno(3.6).$$
\titleb{Photo-hadron production}
The photo--hadron production process is dominated by
photo--pion production, {\it e.g.} for protons at threshold the channels
$\rm p+\gamma \to \pi ^o +p$ and $\rm p+\gamma \to \pi ^+ +n$ dominate
and at higher energies by multipion production. Baryon production
and K-meson production can be neglected in most astrophysical applications.
Pion production
has been treated in detail by
Stecker (1968), Biermann and Strittmatter (1987),
Sikora  {\it et al.} (1987),
Mannheim and Biermann (1989) and Begelman {\it et al.} (1990); see also
references in
Berezinsky {\it et al.} (1990).
Photo--pion production by nuclei of mass number $A$
obeys the simple Glauber rule $\sigma_A\simeq A^{2/3}
\sigma_{\rm p}$. However, in astrophysical circumstances nuclei can not be
accelerated up to
energies above pion threshold, since they are destroyed before by
photodisintegration, see Sect. 3.3.
The latter process has a much lower threshold of about $10$~MeV in the nucleus
rest frame,
whereas pion production requires at least $145$~MeV.

{}For a \kr proton of
Lorentz factor $\gamma_{\rm p}$ traversing an isotropic photon field of number
density
$n_\gamma(\varepsilon )$ one obtains the energy loss rate
$$\eqalign{&-\left({dE_{\rm p} \over {dt}}\right)_\pi^{(\rm p\gamma)}={m_{\rm
p}c^3\over
{2\gamma_{\rm p} }}\cr &\cdot
\int _{\varepsilon _{\rm th}^\prime /(2\gamma_{\rm p} )}
^{\infty }d\varepsilon n_\gamma(\varepsilon )\varepsilon ^{-2}
\int _{\varepsilon _{\rm th}^\prime }^{2\gamma_{\rm p} \varepsilon
}d\varepsilon ^\prime
\varepsilon ^\prime \sigma (\varepsilon ^\prime ){\rm K_{\rm p}}(\varepsilon
^\prime )
\cr}\eqno (3.7)
,$$
where $\sigma (\varepsilon ^\prime )$ and $\rm K_{\rm p}(\varepsilon ^\prime )$
are the total
photo--hadron production cross section and inelasticity, respectively, as a
function of the photon energy in the proton rest frame. The proton rest frame
threshold energy $\varepsilon _{\rm th}^\prime $ is given by Eq.(2.12) for the
respective reaction. Crucial for further evaluation is the knowledge
of both the photo--hadron inelasticity and the cross section as a function
of energy $\varepsilon ^\prime $.
There exist a vast number of
phenomenological particle physics models for the latter two whose
free parameters are adjusted by available accelerator studies of
individual reactions. Stecker (1968) has used an interpolation
to the available experimental photo--hadron production studies to
infer the energy dependence of $\sigma (\varepsilon ^\prime )$ and
$\rm K_{\rm p}(\varepsilon ^\prime )$. Eq.(3.5) was then numerically integrated
to calculate the energy loss rate of protons for photo--hadron
production in the microwave background radiation field and
to compare this with the energy loss rate for pair production
and redshift expansion.
These two loss processes are only significant
at total \kr energies above $10^{18}$eV in this radiation field.
However, in the intense radiation fields of AGN photoproduction of
secondaries is important at much lower energies as
pointed out by
Biermann and Strittmatter (1987) and
Sikora {\it et al.} (1987).
In this context, photo-hadron production was reexamined by various
authors.  Mannheim and Biermann (1989)
and Begelman {\it et al.} (1990) used
the compilation of data by Genzel, Joos and Pfeil (1974).
{}For interactions at high CMS--energies they used
interpolation formulae suggested by specific
assumptions about the gluon content of the
photon structure function.

The
resulting energy loss rate in a thermal radiation field with photons of energy
$\langle kT\rangle$ is given by
$$\left(dE_{\rm p}\over dt\right)_\pi^{\rm (p\gamma)}=
1.8\cdot 10^{10}\left(u_\gamma\over {\rm erg\thinspace cm^{-3}}\right)
\left(\langle kT\rangle\over {\rm eV}\right)^{-2}\ {\rm eV\over s}\eqno(3.8)$$
at threshold
and for all proton energies $E_{\rm p}$
in a nonthermal photon field with spectrum
$n_\gamma(\varepsilon)\propto \varepsilon^{-2}$ by
$$\left(dE_{\rm p}\over dt\right)_\pi^{\rm (p\gamma)}=
4.3\cdot 10^{-7}\left(u_\gamma\over {\rm erg\thinspace cm^{-3}}\right)
\left( E_{\rm p}\over {\rm GeV}\right)^{2}\ {\rm eV\over s}\eqno(3.9).$$
Comparing with Eqs. (3.5) and (3.6) we find that
photo--hadron production, albeit having a cross section
of only $\alpha_{\rm f}\sigma_{\rm pp}$,
can be regarded as an energy loss
process as important as photo--pair production with cross section
$\alpha_{\rm f}\sigma_{\rm T}$.
This is because the former process has
an inelasticity of $1/4$ at threshold, whereas the latter only reaches
an inelasticity of order $2m_{\rm e}/m_{\rm p}$.
However, photo--pair production has a lower threshold energy
by a factor of $1/145$.
Therefore, (taking also into account the fact that the photo--pair production
cross section rises very slowly) the energy
loss rates in inverse power law photon fields
of slope $-2$, Eqs. (3.6) and (3.9),  are
almost equal.  For flatter photon targets,
such as a thermal spectrum (Eqs. (3.5) and (3.8)), or broken power laws
photo--hadron production can be the dominant cooling mechansim.
\titleb{Photodisintegration}
The photodisintegration rate for a nucleus of mass A with the subsequent
release of i nucleons in an isotropic photon field of number density
$n_\gamma(\varepsilon )$ is given by the expression (Stecker 1969)
$$R_{A,i}={c\over 2}\gamma^{-2}\int _0^{\infty }d\varepsilon
n_\gamma(\varepsilon )\varepsilon ^{-2}\int _0^{2\gamma \varepsilon }
d\varepsilon ^\prime \varepsilon ^\prime \sigma _{A,i}
(\varepsilon ^\prime )\eqno (3.10),$$
where $\gamma $ is the Lorentz factor of the \kr nucleus and
$\varepsilon' $ the photon energy measured in the rest frame of the nucleus.
For some nuclei, for example Helium,
the dominant photodisintegration process is proton emission rather than neutron
emission.
Puget {\it et al.} (1976) established that a Gauss\-ian approximation to the
cross section is useful both as an adequate fit to the cross section data
and as an expedient in performing the numerical integration in Eq.(3.10).
These authors calculated the photodisintegration rates for nuclei up
to $A=56$ in a radiation field characteristic for the intergalactic
medium being dominated by the 2.7~K microwave background radiation and
the infrared radiation background. Since for the latter only upper
limits existed in 1976 they chose two power law representations
for the latter which can be considered as giving reasonable lower
(LIR) and upper (HIR) limits on the intergalactic
infrared flux.
{}For $\ ^{56}$Fe nuclei the energy loss rates from
pair production and photodisintegration are comparable in the energy
range near $10^{20}$eV, whereas below $10^{19}$eV the photodisintegration
time scale becomes longer than the expansion losses.
Puget {\it et al.}(1976) have also pointed out that for photodisintegration
the average fractional energy loss rate obeys
$$E^{-1}(dE/dt)=A^{-1}(dA/dt)\eqno (3.11),$$
because the nucleon emission is isotropic in the rest system of the nucleus.

The energy needed
to overcome the threshold energies for photodisintegration,
photo--pair and photo--hadron production, respectively,
does not come
from the low energy target photon, but has to be
supplied by the nucleus on expense of its kinetic
energy as measured in the lab frame.  Thus, because the
threshold energies for photodisintegration and photo-pair production
are negligible compared to the nucleus rest mass,
the Lorentz factor of the nucleus is
only marginally lowered in these processes.  The contrary is true
in the case of photo-hadron production where the Lorentz factor
is lowered significantly.

Another noteworthy difference concerns the (A/Z) dependence of
the loss rates. The energy loss time for pair production calculated from
Eq.(3.1) varies as $t _{\rm pair}\propto AZ^{-2}$, whereas the energy loss time
from photodisintegration implied by Eq.(3.11) is almost independent from A.
Thus the importance of pair production as an energy loss process
relative to photodisintegration losses decreases with decreasing Z
when we consider \kr nuclei lighter than iron.
\titlea{Interactions between \kr \nuci and matter}
As important as particle--photon collisions are inelastic reactions of
\kr \nuci with atoms and molecules of interstellar and intergalactic
matter. In inelastic p--p, p--$\alpha $, and $\alpha $--p collisions mainly
charged and neutral pions are produced. The charged pions
$\pi ^{\pm }$ decay into muons and (anti-) neutrinos, and the muons
decay into electrons and (anti-) neutrinos.
These decays are regarded as the most important
secondary source of energetic cosmic electrons, positrons and neutrinos.
Neutral pions $\pi^\circ$ decay after a mean lifetime of $9\cdot 10^{-17}$s
into two high--energy gamma rays.

Also of interest, in particular for the origin of nuclear gamma ray
lines, is the excitation of nuclei by collisions between fast \kr particles
and nuclei of the interstellar medium. A discussion of these interactions
is given after the study of pion production.

Secondary neutrons and protons can also be produced in reactions
not involving pions. Such processes have importance in high--energy
processes on the sun (Murphy {\it et al.} 1987) but will not be discussed
here.

\titleb{Gamma--ray, electron, positron and neutrino source
functions}
Assuming that the decaying particles
are distributed isotropically in their rest frame we can relate the energy
spectrum of the decay products to the energy spectrum of the decaying
particles\fonote{{}For two--body
and three--body decays these relations are
derived in Ch.7.6 of Hagedorn (1973).}.
The omnidirectional (i.e. integrated
over the whole solid angle) differential gamma ray source function
$q_\gamma(E_{\gamma })$
is related to the omnidirectional differential neutral pion source function
$q_{\pi^o}(E_\pi)$
according to
$$\eqalignno{&q_{\pi ^o}(E_{\gamma })=& (4.1)\cr &
\cdot 2\int_{E_{\gamma }+[(m_{\pi }c^2)^2/(4E_{\gamma }
)]}^{\infty }dE_{\pi }q_{\pi ^o}(E_{\pi })
\left[E_{\pi }^2\ -\ m_{\pi }^2c^4\right]^{-1/2}\cr}.$$
The decay of the charged pions is more complicated and has been treated
in detail by Scanlon and Milford (1965). Let us first consider the
decay $\pi ^{\pm }\to \mu ^{\pm }+\nu _{\mu }
(\bar \nu _{\mu })$. The Lorentz factor
$\gamma _{\mu }^\prime $ of the muon in the rest frame of the pion
follows readily from the conservation of four momentum in the
2--body decay $P_{\pi }=P_{\mu }+P_{\nu }$ implying with
$m_{\nu }=0$ that $P_{\pi }P_{\mu }=[m_{\pi }^2+m_{\mu }^2]c^2/2$.
In the rest frame of the pion
$P_{\pi }P_{\mu }=m_{\pi }m_{\mu }c^2\gamma _{\mu }^\prime$ implying
$$\gamma _{\mu }^\prime ={m^2_{\pi }+m^2_{\mu }\over
{2m_{\pi }m_{\mu }}}\eqno (4.2).$$
With $m_{\pi }c^2\simeq 140$~MeV and $m_{\mu }c^2=106$~MeV we obtain
$\gamma _{\mu }^\prime =1.04$. This low value implies that, as long
as the velocity of the pion in the laboratory frame is not
exceedingly small ($\gamma _{\pi }>1.04$), the muon can be treated
as essentially at rest in the rest frame of the pion. Therefore,
in the laboratory frame, the muon moves with almost the same speed
as the pion, so that per unit Lorentz factor, the muon and pion
source functions are equal (Ramaty 1974), hence
$$q_{\mu^\pm }(E _{\mu })={m_\pi\over m_\mu}  q_{\pi ^\pm}(E _{\pi })\eqno
(4.3).$$
The decay $\mu ^\pm \to e^\pm +\nu _{\rm e}(\bar \nu _{\rm e})
+\bar \nu _{\mu }(\nu _{\mu })$ is a 3--body
process. The relation of the secondary electron and positron production
spectrum to the muon source function has been calculated by
Scanlon and Milford (1965), Ramaty \& Lingenfelter (1966),
Perola {\it et al.} (1967) and Ramaty (1974);
their approximate formula is
$$q_{\rm e^\pm}(E_{\rm e^\pm})= {m_\pi\over 70m_{\rm e}} q_{\pi^\pm}
\left(E_\pi\over 70\right)\eqno(4.4).$$
The corresponding relation of the neutrino production spectra
to the muon source function has been given by Zatsepin and Kuzmin
(1962) and Marscher {\it et al.} (1980).

The mean energies of the produced secondaries in the laboratory frame are given
by
$\langle E_{\rm e^\pm}\rangle={1\over 4}E_{\pi^\pm}$, $\langle
E_{\nu,\bar\nu}\rangle
={1\over 4}E_{\pi^\pm}$ and
$\langle E_\gamma\rangle={1\over 2}E_{\pi^\circ}$.
Therefore, using isospin symmetry
${1\over 3}L_\pi=L_{\pi^+}=L_{\pi^-}=L_{\pi^\circ}$
we obtain the  luminosities
$L_\gamma=L_{\nu,\bar\nu}={1\over 2}L_\pi$ of the produced secondaries.
The gamma ray luminosity includes account of the gamma rays produced
by neutral pion decay $L_\gamma(\pi^{\circ})=L_{\pi^\circ}$ and of
the gamma rays produced by rapid cooling of the produced pairs
from charged pion decay $L_\gamma(e^\pm\rightarrow \gamma)={1\over 6}
L_\pi$.
\nobreak
\titleb{Pion production spectra}
To obtain the pion source spectra required in Eqs. (4.1), (4.3) and (4.4)
we start from the respective
pion power of a single relativistic proton
of total energy $E_{\rm p}=\gamma _{\rm p}m_{\rm p}c^2$
$$\eqalign{&P(E_\pi,E_{\rm p})=1.3cE_\pi n_{\rm H} \cr &
\cdot\xi(E_{\rm p})\sigma_{\rm pp}^\pi
\delta[E_\pi-\langle E_\pi\rangle]H[E_{\rm p}-E_{\rm th}]\cr}\eqno(4.5),$$
where $n_{\rm H}=n_{\rm HI}+n_{\rm HII}+2n_{\rm H_2}$
denotes the target hydrogen density, $\xi$ the multiplicity of the pions and
$\langle E_\pi\rangle$ the average energy of a single produced pion.
H denotes the Heaviside step--function ($H(x)=1$ for $x\ge 0$,
and $H(x)=0$ for $x<0$), $E_{\rm th}=1.22$~GeV the threshold energy for pion
production
and $\sigma_{\rm pp}^\pi\simeq \sigma_{\rm p,inel}\simeq 3\cdot
10^{-26}$~cm$^2$
the total pion cross section.  The factor $1.3$ accounts for the chemical
composition of
the interstellar matter.  The error introduced by assuming a constant cross
section
grows only logarithmically with energy.

The number of pions produced in each hadronic interaction increases rather
slowly.  Up to laboratory proton energies of $10^4$~GeV the increase can be
approximated
by a power law, {\it i.e.} for charged pions
$$\xi_{\pi^\pm} (E_{\rm p})\simeq 2 \left[(E_{\rm p}-E_{\rm th})/{\rm
GeV}\right]^{1/4}\eqno(4.6).$$
Isospin symmetry yields the multiplicity of neutral pions $\xi_{\pi^o}={1\over
2}\xi_{\pi^\pm}$.
This is in agreement
with Fermi's original theory (Fermi 1950)
of pion
production, in which a thermal equilibrium of the resulting pion cloud
in the CMS
is assumed.  However,
at still higher energies, Eq.(4.6) overestimates the number of pions.

Since the limiting value of the inelasticity $K_{\rm p}$
(cf. Section 3.2) is roughly 1/2 (on average,
a leading nucleon and a pion cloud each carrying half of the
total incident energy leaves the interaction fireball),
the energy dependence of the mean
pion energy below $10^4$~GeV is given by
$$\langle {E_{\pi }}\rangle(E_{\rm p})=K_{\rm p}(E_{\rm p}-m_{\rm p}c^2)/\xi
\simeq {1\over 6}\left(T_{\rm p}\over {\rm GeV}\right)^{3/4}\ {\rm GeV}
\eqno(4.7),$$
where $T_{\rm p}$ denotes the kinetic energy of the proton
(cf. Cheng 1972, Stecker 1973).
{}For a differential
number density distribution of \kr protons
$n_{\rm p}(E_{\rm p})$  we obtain for the
pion source functions
$$q_{\pi ^o,\pi ^\pm }(E_{\pi })=
{1.26\over {E_{\pi }}}\int_1^\infty dE_{\rm p}
n_{\rm p}(E _{\rm p})P_{\pi ^o,\pi ^\pm } (E_{\pi },E_{\rm p})\eqno (4.8),$$
where the factor 1.26 accounts for the contribution of $\alpha $--p,
$\alpha -\alpha $ collisions
and collisions of higher metallicity {\krs}.
It
is derived using the ratio of the
respective inclusive cross section to the p--p inclusive cross
section and the known elemental composition of the \krs .

{}For the astrophysically important case of a power
law distribution of \kr protons and $\alpha $--particles
$$n_{\rm p}(E_{\rm p})=
n_{\rm p,\circ}\left[(E_{\rm p}-m_{\rm p}c^2)/{\rm GeV}\right]^{-s}\eqno(4.9)$$
the source functions reduce to
$$\eqalign{&q_{\pi ^o}(E_{\pi })=q_{\pi ^+}(E_{\pi })=q_{\pi ^-}(E_{\pi
})\simeq
13.1 c n_{\rm p,\circ}\cr &\cdot n_{\rm H}\sigma_{\rm p,inel}
\left[6(E_\pi/ {\rm GeV})\right]^{-{4\over 3}(s-{1\over 2})}\cr}\eqno(4.10).$$
{}For the special case $s=2$ the slope of the pion spectra is the same as the
generating proton spectrum, whereas for $s>2$ the pion spectra steepen due to
multiple pion production.  Above $10^4$~GeV the energy dependence of
multiplicity and average single pion energy relax, so that the power law index
approaches $s$ again.
{}For an accurate treatment of pion production near threshold see Dermer
(1986).
\titleb{Energy loss by pion production}
Integrating the pion power (4.5) over all pion energies $E_{\pi }$
we obtain the energy loss of a single relativistic proton or neutron
due to
pion production
$$\eqalignno{&-\left({dE_{\rm p}\over{dt}}\right)_{\pi }=
\int_0^{E_{\pi _{\rm max}}}dE_{\pi }P(E_{\pi },E_{\rm p} )&(4.11)\cr &
=0.65cn_{\rm H}\sigma _{\rm p,inel}(E_{\rm p}-m_{\rm p}c^2)
H[E_{\rm p}-1.22\thinspace{\rm GeV}]\cr}.
$$
The way, in which energy is imparted on pions during  hadronic
interactions, is irrelevant to the total energy loss, which
is governed solely by the inelasticity and total pion cross section.

The case of nuclei with $A>1$ scattering on the interstellar matter involves
both breakup of
the nucleus and pion production and is treated in Section 4.6.
\titleb{Excitation of nuclei}
Excited nuclei produced by collisions between fast particles and nuclei
undergo nuclear transitions to their ground states whereby gamma rays
may be emitted. The excited states of nuclei that are appreciably abundant
in stellar and interstellar matter, i.e. the nuclei C, N, O, Ne, Mg, Si,
and S emit the most important gamma ray lines
(Meneguzzi and Reeves 1975).

Two types of collisions occur :(i) collision of fast
\kr nuclei with p
and $\alpha$--particles of the stellar or interstellar target gas
at rest, and (ii) collision of fast p and $\alpha $--particles
of the cosmic radiation with heavier target nuclei at rest.
Summing over all inelastic cross sections and the appropriate lifetimes
of the excited states for the various nuclei the emergent nuclear
gamma ray line intensities have been calculated in detail for
different energy spectra of the energetic \kr nuclei in astrophysical sites as
solar and stellar flares, galactic centers and supernova envelopes
({\it e.g.}, Ramaty {\it et al.} 1979).
The resulting gamma ray intensities exhibit a great
wealth of spectral structure, ranging from very narrow to broad features,
depending on the composition and energy spectrum of the energetic particles,
 and the composition and state of the ambient medium.

With the emergent intensity being very sensitive to the assumed state of
the target material and the spectrum of incoming \kr nuclei it is
clear that the succesful spectroscopy of these lines will provide many
detailed clues on the ongoing astrophysical processes. The contribution
as an energy loss process of \kr nuclei, however, is negligibly small
as compared to the Coulomb and ionization losses which we are going to discuss
next.
\titleb{Coulomb and ionization interactions}
\titlec{Interactions in fully ionized plasma}
The rate of energy loss of a fast test ion
of mass $M=Am_{\rm p}$, charge Ze and
velocity v=$\beta $c
in a completely ionized thermal plasma
consisting of s species of respective mass $m_{\rm s}$, charge $Z_{\rm s}e$,
density
$n_{\rm s}$ and temperature $T_{\rm s}$
density $n_{\rm e}$ has been calculated by
Butler and Buckingham (1962), Sivukhin (1965) and Gould (1972) as
$$-\left({dE\over{dt}}\right)_{\rm coul}={{4\pi Z^2e^4\ln \lambda }\over
v}\sum_{\rm s}
{Z_{\rm s}^2n_{\rm s}\over {m_{\rm s}}}W_{\rm s}\left(
{\beta \over {\beta _{\rm s}}}\right)$$
$$={3\ c\ \sigma _T\ m_{\rm e}c^2Z^2\over {2\beta }}\sum_{\rm s}
{m_{\rm e}\over {m_{\rm s}}}Z_{\rm s}^2n_{\rm s}\ W_{\rm s}\left({\beta \over
{\beta _{\rm s}}}\right)
\eqno (4.12),$$
where
$$\beta _{\rm s}\equiv \left({2kT_{\rm s}\over {m_{\rm
s}c^2}}\right)^{1/2}\eqno (4.13),$$
and
$$\eqalignno{&W_{\rm s}(x)\equiv &(4.14)\cr &
{2\over {\pi ^{1/2}}}\left[\int_0^xdy\exp(-y^2)-\left(1+{{m_{\rm s}}\over
M}\right)
x\exp(-x^2)\right]\cr}.$$
The Coulomb logarithm appearing in Eq.(4.12) $\ln \lambda $
is equal to 20 for a wide range of plasma densities and temperatures
(see Sivukhin 1965).
Because of the small electron--proton mass ratio $m_{\rm e}/m_{\rm p}=1/1836$
it is easy to see that for all values of the metallicity of the plasma
the Coulomb collisions are dominated by scattering off the thermal
electrons, i.e.
$$\sum_{\rm s}{m_{\rm e}\over {m_{\rm s}}}Z_{\rm s}^2n_{\rm s}W_{\rm s}\simeq
n_{\rm e}W_{\rm e}\eqno (4.15),$$
so that Eq.(4.12) reduces to
$$-\left({dE\over {dt}}\right)_{\rm coul}\simeq {{30 c \sigma _Tm_{\rm
e}c^2Z^2}\over {\beta }}
n_{\rm e}\ W_{\rm e}\left({\beta \over {\beta _{\rm e}}}\right)\eqno (4.16),$$
where numerically
$$\beta _{\rm e}=0.026(T_{\rm e}/2\cdot 10^6\thinspace {\rm K})^{1/2}
\eqno (4.17).$$

{}For small and large arguments we may approximate Eq. (4.14) as
$$\eqalign{&W_{\rm e}\left({\beta \over {\beta _{\rm e}}}\right)
\simeq \cr &
\left\{\eqalign{{2\over {\pi ^{1/2}}}{\beta \over {\beta _{\rm e}}}
\left[-{m_{\rm e}\over M}+\ ({2\over 3}+{m_{\rm e}\over M})
({\beta \over {\beta _{\rm e}}})^2\right]\ &\hbox{\rm for}\ \beta \ll \beta
_{\rm e} \cr
1\ &\hbox{\rm for}\ \beta \gg \beta _{\rm e} \cr}\right\}\cr }\eqno (4.18).$$
{}For values of $\beta \le \beta _{\rm c}(A,T_{\rm e})$ where
$$\beta _{\rm c}(A,T_{\rm e})\equiv [1.5m_{\rm e}/(Am_{\rm p})]^{1/2}\beta
_{\rm e}
=0.0286\ A^{-1/2}\beta _{\rm e}\eqno (4.19)$$
we find $W_{\rm e}<0$, indicating that at low particle velocities
Coulomb collisions accelerate ($(dE/dt)>0$) the test ion up to velocity
$\beta _{\rm c}(A,T_{\rm e})$.
In the intermediate range $\beta _{\rm c}(A,T_{\rm e})\le \beta \le \beta _{\rm
e}$ we obtain
according to Eq.(4.18) $W_{\rm e}\simeq (4/3)\pi ^{-1/2}(\beta /\beta _{\rm
e})^3$
and beyond $\beta \ge \beta _{\rm e}$, $W_{\rm e}$ quickly approaches the
asymptotic
value $W_{\rm e}=1$. The interpolation formula
$$W_{\rm e}(\beta \ge \beta _{\rm c}(A,T_{\rm e}))={{\beta ^3}\over{x_m^3+\beta
^3}}
\eqno (4.20),$$
with
$$\eqalign{&x_m\equiv \left({{3\pi ^{1/2}}\over 4}\right)^{1/3}\beta _{\rm
e}\cr & =1.10\beta _{\rm e}
=0.0286(T_{\rm e}/2\cdot 10^6\thinspace{\rm K})^{1/2}\cr}\eqno (4.21)$$
being independent of A, provides an excellent fit to the exact variation
over the whole range $\beta \ge \beta _{\rm c}$.

Using the interpolation formula (4.20) in Eq.(4.16) we obtain
$$-\left({dE\over {dt}}\right)_{\rm coul}\simeq 3.1\cdot 10^{-7}Z^2\
\left(n_{\rm e}\over {\rm cm^{-3}}\right)
{\beta ^2\over {x_m^3+\beta ^3}}\ {\rm eV\over s}\eqno (4.22).$$
Below the
Coulomb barrier
$$T_{\rm m}
\simeq 0.61 \left( T_{\rm e}/2\cdot
10^6\thinspace {\rm K}\right)\ {\rm MeV}\eqno(4.23)$$
the total energy loss rate varies linearly proportional to
the kinetic energy $T_{\rm p}$,
whereas beyond the barrier it decreases $\propto T_{\rm p}^{-1/2}$ for
$T_{\rm p}\ll m_{\rm p}c^2$.
\titlec{Interactions in neutral matter}
The energy loss of particles traversing neutral matter was first
calculated by Bohr using the classical theory. In quantum mechanics
the problem has been investigated by several authors including
Moeller, Bethe, Williams and Bloch in a satisfactory way using
various atomic models. We follow the treatment of Heitler (1954);
a detailed summary of the results can also be found in Hayakawa
(1969, his Ch.2.2).

The energy loss rate of an ion of charge $Z$ and mass $M=Am_{\rm p}$ traversing
neutral
matter consisting of s sorts of atoms and molecules with concentrations
$n_{\rm s}$ and charges $z_{\rm s}$ is ( Heitler 1954, Ch.37;
Evans 1955, Ch.18; Northcliffe 1963)
$$\eqalign{&-\left({dE\over {dt}}\right)_{\rm ion}\left(\beta \ge \beta
_\circ\right)
=\cr &
{3\ c\sigma _T\ Z^2\ m_{\rm e}c^2\over {4\beta }}\sum_{\rm s}n_{\rm s}[B_{\rm
s}+
\ B^\prime (\alpha_{\rm f} Z/\beta )]\cr }
\eqno (4.24),$$
with
$$B_{\rm s}= \left[\ln\left({2m_{\rm e}c^2\beta ^2Q_{max}\over {I_{\rm
s}^2(1-\beta ^2)}}\right)-2\beta ^2-
{2C_{\rm s}\over {z_{\rm s}}}-\ \delta _{\rm s}\right]\eqno (4.25).$$
The largest possible
energy transfer from the incident particle
to an initially stationary atomic electron follows from the conservation laws
for momentum and energy
and is
$$\eqalign{Q_{max}=&{T+2Mc^2\over
{1+[(M+m_{\rm e})^2c^2/(2m_{\rm e}T)]}}\cr &\simeq
{2m_{\rm e}c^2\beta ^2\gamma ^2\over {1+[2m_{\rm e}T/(m^2c^2)]}}
\cr}\eqno (4.26),$$
since $M\gg m_{\rm e}$.  $T=E-Mc^2$ denotes the kinetic energy of the ion.
The function
$$B^\prime (x)=2[\Psi (0)-\ \Re \Psi (\imath x)]\eqno (4.27),$$
where $\Psi (x)$ denotes the Digamma function, is important for slow
particles or particles with large atomic number so that the product
$\alpha_{\rm f} Z/\beta =Z/(137\beta )$ is no longer small to unity. It is
straightforward to show that for small values of x$\ll 1$, $B^\prime (x\ll 1)
\to 0$, whereas in the opposite case, $B^\prime (x\gg 1)\to -2\ln x$.
$I_{\rm s}$ denotes the geometrical mean of all ionization and excitation
potential
of the absorbing atom or molecule s and is defined by
$$z_{\rm s}\ln I_{\rm s}\equiv \sum_{n,l}f_{n,l}^sA_{n,l}^s$$
where $f_{n,l}^s$ is the sum of the oscillator strengths for all optical
transitions of the electron in the n,l shell of the atom s and is close to
unity, while $A_{n,l}^s$ is the mean excitation energy of the n,l shell and
be put equal to the ionization potential with sufficient accuracy. Empirical
values for $I_{\rm s}$ can be determined for each value of $z_{\rm s}$. {}For
most
elements $I_{\rm s}$ in eV is roughly $13z_{\rm s}$; for the for
astronomical applications important very light elements hydrogen and helium
the experimental values are $I_H=19$~eV and $I_{He}=44$~eV, respectively.
The shell--correction term $(C_{\rm s}/z_{\rm s})$ and the density correction
term $\delta _{\rm s}$ of Sternheimer (1952), resembling
the effect of the polarization of the medium on
the ionization loss, lead to a reduction of the
logarithmic $\ln \gamma $ increase of the ionization rate at relativistic
energies to a constant value.
Equation (4.24) can be used for particle velocities
high compared to the characteristic velocity of the medium's electrons.
In the case of atomic hydrogen this characteristic velocity is
determined by the orbital velocity of the electrons
$\beta _\circ=1.4e^2(\hbar c)^{-1}=0.01$ corresponding to a kinetic energy of
$$T_\circ=49A\ \rm keV\eqno (4.28).$$
{}From the asymptotic behaviour of the function $B^\prime $
one deduces immediately that
$B_{\rm s}+B^\prime \simeq B_{\rm s}$ for $\beta >max[\beta _\circ,Z/137]$,
whereas in the case
$\beta _\circ<\beta <Z/137$ we obtain Bohr's original classical formula
$$B_{\rm s}+B^\prime \simeq 2\ln {2c\hbar m_{\rm e}c^2\beta ^3\over {I_{\rm
s}Ze^2}}\eqno (4.29).$$

Below the kinetic energy $T_\circ$ the ionization rate changes.
This has been clearly
established by laboratory measurements of the stopping cross section
per atom as collected by Whaling (1958).
{}For ions of velocity large compared to $\beta _\circ$
there is excellent agreement between formula (4.24) and the experimental
values. However, for ions of lower velocity there is according to Whaling
(1958) no regularity in the energy dependence of the stopping cross section
for all absorbing atoms. Hayakawa (1969) discusses some relevant
physical processes at these low ion velocities.
Ginzburg and Syrovatskii (1964)
give a useful ionization loss rate for slow ion velocities
for the case that the ion atomic number Z lies between
${1\over 4}Z\le z_{\rm s}\le 4Z$,
$$\eqalign{&-\left({dE\over {dt}}\right)_{\rm ion}(\beta \le \beta
_\circ)=2.11\cdot 10^{-2}
\beta ^2\cr &\cdot\sum_{\rm s}
\left(n_{\rm s}\over{\rm cm^{-3}}\right)(Z+z_{\rm s})\ {\rm eV\over s}\cr}\eqno
(4.30).$$
which agrees well with the experimental data of Whaling (1958).
So again, as in the case of Coulomb losses, the ionization losses
have a maximum at $T_\circ$ given in Eq.(4.28) below which the loss
rate varies proportional to $\propto T$ while it varies
$\propto T^{-1/2}$ at non-relativistic energies above $T_\circ$
according to Eq.(4.24). Using the asymptotic behaviour of
$Q_{max}$ from Eq.(4.26) $Q_{max}(T\le 918A^2m_{\rm p}c^2)
\simeq 2m_{\rm e}(c\beta \gamma )^2$
and $Q_{max}(T\ge 918A^2m_{\rm p}c^2)\simeq T$ as
well as the values of the ionization
potentials of hydogen and helium, taking into account the chemical composition
of the interstellar medium we obtain for Eq.(4.24)
$$\eqalignno{&-\left({dE\over {dt}}\right)_{\rm ion}={3\over 2}
\sigma _T\ c\ Z^2m_{\rm e}c^2\sum_{\rm s}n_{\rm s}&(4.31)\cr &
\cdot \left\{\eqalign{
\ln {2m_{\rm e}c^2\over {I_{\rm s}}}+\ \ln \beta + {\beta ^4\over 2}
\ &\hbox{\rm for}\ T_\circ\le T\le 918 A^2m_{\rm p}c^2 \cr
\ln {2m_{\rm e}c^2\over {I_{\rm s}}}+\ 3.411+\ 0.5\ln A
\ &\hbox{\rm for}\ T\ge 918 A^2m_{\rm p}c^2 \cr}\right\}\cr &
\simeq 1.82\cdot 10^{-7}{\beta }^{-1}Z^2([n_{\rm HI}+\ 2n_{{\rm H_2}}]/{\rm
cm^{-3}})
\cr &
\cdot \left\{\eqalign{
1+\ 0.0185\ln \beta \ &\hbox{\rm for}\ T_\circ\le T\le 918A^2m_{\rm p}c^2 \cr
1.315[1+\ 0.035\ln A]\ &\hbox{\rm for}\ T\ge 918A^2m_{\rm p}c^2 \cr}\right\}
\ {\rm eV\over s}\cr}.$$
At very low energies, $T\le T_\circ$, Eq.(4.30) has to be used.

{}For \kr protons a useful interpolation formula for all energies
below $T\le 918m_{\rm p}c^2$ is
$$\eqalign{&-\left({dE\over {dt}}\right)_{\rm ion,Z=1}=1.82\cdot
10^{-7}([n_{\rm HI}+\
2n_{{\rm H_2}}]/{\rm cm^{-3}})\cr &
\bigl(1+\ 0.0185\ln \beta H[\beta - \beta _\circ]\bigr)
{2\beta ^2\over {\beta _\circ^3+\ 2\beta ^3}}\ {\rm eV\over s}\cr}\eqno
(4.32)$$
which attains its maximum at $\beta _\circ=0.01$ corresponding to
the energy $T_\circ$ given in Eq.(4.28). The function H in (4.32) denotes
the Heaviside step function.

The ionization loss for a higher metallicity ion of charge Z
can be expressed in terms of the ionization loss of protons, Eq.(4.36),
as (Brown and Moak 1972)
$$\left({dE_N\over {dt}}\right)_{\rm ion}=Z_{\rm eff}^2\left({dE_N\over
{dt}}\right)_{\rm ion,Z=1}
\eqno (4.33)$$
at the same kinetic energy per nucleon $T=E-Mc^2$ with the effective charge
$$Z_{\rm eff}=Z\left(1-1.034\exp{[-137\beta Z^{-0.688}]}\right)\eqno (4.34),$$
where $\beta $ is the ion's velocity in units of c. The effective charge
at small energies is less than $Z$ since complete stripping of the
heaviest ions occurs only at energies above 100~MeV/nuc.

If the primary particle is an electron or a positron, a modification
ought to be made in Eq.(4.24) reflecting the fact that for large
energy transfer exchange and spin effect play some role.
\titleb{Catastrophic losses from fragmentation
and radioactive decay}
In inelastic proton--nucleus and $\alpha $--nucleus
collisions with
atoms and molecules of the interstellar and intergalactic
target gas \kr \nuci with charge greater than 1
can fragment in reactions $\rm A+\ (p,\alpha )\to (A-k)+$anything, with
k=1,2,3,...(A-1). The individual fragmentation cross sections have
been tabulated by Silberberg and Tsao (1973, 1990). Letaw {\it et al.} (1983)
have derived a very useful empirical formula for the total inelastic
cross section of protons on nuclei with $A>1$. This cross section
covers both, pion production and fragmentation.  At high kinetic energies ($T
\ge 2$~GeV) the formula reproduces experimental data to within reported
errors ($\le 2 \% $) whereas at lower energies
the maximum error is 20 $\% $ at 40~MeV.
According to Letaw {\it et al.} (1983) (1 mb=$10^{-27}\ {\rm cm}^2$) at high
energies
$$\eqalign{&\sigma _{\rm tot}(T\ge 2{\rm GeV})=\cr &
45\ A^{0.7}[1+\ 0.016\sin(1.3-\ 2.63
\ln A)] \ {\rm mb}\cr}\eqno (4.35)$$
is independent of energy and increases with increasing \nuc mass number A.
\begfig8cm
\figure{1}{{\bf Upper panel:}  Total energy loss rate ${1\over n_{\rm H}}
{dE_{\rm p}\over dt }$
of \kr protons
in the local
interstellar medium ({\it solid curve}) due to the various loss processes
({\it dashed curves}).
The neutral gas density was assumed to be $n_{\rm HI}+\ 2n_{{\rm H_2}}$
=1.14~cm$^{-3}$ and the density of thermal electrons was assumed
to be $n_{\rm e}=0.01$~
cm$^{-3}$ at a temperature $T_{\rm e}$= 3$\cdot 10^5$~K.
We demonstrate Bethe-Heitler pair production
losses by assuming a thermal radiation field with luminosity
$L_{\rm ir}=10^{44}$~erg/s
and mean photon energy $kT=1$~eV.  The onset of pair production
is seen above the threshold energy $E_{\rm p, th}^{(\rm e^\pm)}\simeq 6\cdot
10^8$~MeV, whereas photo--pion production occurs at still higher energies.
The
{\it dotted curve} shows the effect of an enhanced
density $n_{\rm HI}+\ 2n_{{\rm H_2}}=100$~cm$^{-3}$ which leads to an
increasing importance of $pp$-collisions relative to the escape losses.
{\bf Lower panel:}  Corresponding stationary \kr distribution
$n_{\rm p}$ multiplied
by $T ^{2}$ in arbitrary units.  The {\it dotted curve} demonstrates the
effect of an enhanced neutral gas density which is to flatten the \kr
spectrum in the GeV range.}
\endfig
At energies below 2~GeV the total inelastic cross section is not independent
of energy. In general terms, it decreases to a minimum ($\sim 15\% $ below
the high energy value) at 200~MeV. It then sharply increases to a maximum
at $\sim 20$~MeV (60\% above the high energy value). Below this energy
resonance effects become dominant and the cross section fluctuates
rapidly. In the low--energy range the approximation is
$$\eqalign{&\sigma_{\rm tot}(T \le 2\ \hbox{\rm GeV})=
\sigma_{\rm tot}(T\ge 2\ \hbox{\rm GeV})\cr &\cdot
[1- \ 0.62\exp (-T/200)\sin (10.9\cdot
T^{-0.28})]\cr}\eqno(4.36),$$
where $T$ is in MeV.

Fragmentation loss times for \kr \nuci follow immediately from the total
inelastic cross sections as
$$\eqalign{&t _{\rm f}(T)=\left[\sum_{\rm s}n_{\rm s} c\beta
\sigma _{\rm tot}(T)\right]^{-1}
\cr &=
(1.3n_{\rm H}c\beta \sigma _{\rm tot}(T)\ )^{-1}\cr }\eqno (4.37).$$
At relativistic energies we obtain an energy--independent fragmentation
lifetime of
$$\eqalign{&t _{\rm f}(T>2 {\rm GeV})\simeq 2\cdot 10^7 \ A^{-0.7}\cr &\cdot
(n_{\rm H}/{\rm cm^{-3}})^{-1}\
\rm yrs\cr}\eqno (4.38).$$
The fragmentation losses are catastrophic losses, i.e. they do not conserve
the total particle number of the considered nucleus. {}For this reason
we have to describe the loss process by a loss time rather than
an energy loss rate which is appropriate for continuous losses
that redistribute the particles in energy space but conserve the total
number of particles.

In case of unstable particles (as $Be^{10}, \ Al^{26}$) the total
catastrophic loss time $t _{\rm c}$ is determined by
$$t _{\rm c}^{-1}=t _{\rm f}^{-1} +\ (\gamma T_{\rm
decay}^o)^{-1}\eqno(4.39),$$
where $T_{\rm decay}^o$ is the radioactive half--life time at rest.
\titleb{Total energy loss rate from interactions with matter}
Combining the energy loss rates due to pion production Eq.(4.11),
Coulomb scattering Eq.(4.22) and ionization Eq.(4.32) and introducing
the degree of ionization, $x=n_{\rm e}/n_{\rm H}$ where
$n_{\rm H}=n_{\rm HII}+n_{\rm HI}+2n_{\rm H_2}$, we obtain for the total
energy loss rate of \kr protons (A=1) in interactions with matter
$$\eqalignno{&-\left({dE_{\rm p}\over {dt}}\right)_{\rm tot,Z=1}(\beta)=
1.82\cdot 10^{-7}(n_{\rm H}/{\rm cm^{-3}})\cr &
\cdot (1-x)\Bigl[ [1+(0.0185\ln\beta H[\beta - 0.01])]{2\beta ^2\over {10^{-6}
+2\beta ^3}}
&(4.40)\cr &+\ 1.69x
 H[\beta -7.4\cdot 10^{-4}
(T_{\rm e}/2\cdot 10^6\thinspace{\rm K})^{1/2}]\cr &
\cdot {\beta ^2\over {\beta ^3+2.34\cdot 10^{-5}
(T_{\rm e}/2\cdot 10^6\thinspace {\rm K})^{3/2}}}\cr &
+3[(1-\beta^2)^{-1/2}-1]H[\beta -0.64]
\Bigr]{\rm eV\over s}\cr}.$$
In order to understand the relative importance of the different energy
loss processes in the local interstellar medium we have calculated
the individual loss rates by using parameters as measured near the solar system
in the local galactic neighbourhood. Fig.~1 shows the result. Whereas
Coulomb losses play only a minor role, ionization losses dominate the
total energy loss rate below the kinetic energy $T^*\approx 450$~MeV,
corresponding
to a velocity $\beta ^*\approx 0.74$, whereas pion production losses become
dominant above $T^*$. It should be noted that the value of $T^*$ is
independent of the interstellar gas density, since
both ionization and pion production scale with the gas density.
It is only determined by the ratio of the respective interaction cross
sections. {}For the associated total loss time scale
$t _{\rm tot,Z=1}\equiv T_{\rm p}/[|dE_{\rm p}/dt|]_{\rm tot, Z=1}$ with
$T_{\rm p}=E_{\rm p}
-m_{\rm p}c^2$
we find for $x\ll 1$
$$\eqalignno{&t _{\rm tot,Z=1}
\simeq {1\over {(n_{\rm H}/
{\rm cm^{-3}})}}& \cr &\cdot
\left\{\eqalign{50\ \hbox{\rm yrs}\ \ &\hbox{\rm for}\ T_{\rm p}\approx
T_\circ\approx 45\ \hbox{\rm keV} \cr
5\cdot 10^7\ (T_{\rm p}/T^*)^{1.5}\ \hbox{\rm yrs}\ \ &\hbox{\rm for}\
T_\circ< T_{\rm p}\le T^*\approx 450\ \hbox{\rm MeV} \cr
5\cdot 10^7\ \hbox{\rm yrs}\ \ &\hbox{\rm for}\ T^*< T_{\rm p}\le 10^4\ {\rm
GeV}
\cr}
\right\}\cr}.$$
The total energy loss rate for higher metallicity ions ($Z>2$) in matter
follows
readily from Eq.(4.40) using Eqs.(4.16), Eq.(4.33) and (4.34).
We turn now to the observational
consequences of \kr \nuc interactions.

\titleb{Ionization and heating rate of interstellar matter
by \krs }
As a result of ionization interactions of neutral interstellar matter
with \kr \nuci (Section 4.5.2) the latter becomes ionized.
According to Heitler (1954) the average energy loss is also, with
fairly great accuracy, representative of the average primary ionization
of the particle. The fractional number of cases in which a collision
with an atom results in ionization rather than excitation to a discrete level
(discussed in Section 4.4) is almost independent of the energy of the primary,
and the same is true for the average energy transferred to the ionized
electron. Thus the number of primary ion pairs formed per time interval
is very nearly proportional to the average loss by collision given in
Eq.(4.24). The number of primary ion pairs can be inferred from the fact
that about one ion pair is formed when the primary loses an amount of
32 eV. This figure is practically independent of the nature of the particle
and its energy (at least if the latter is not excessively small).
It should be noted that the secondary particles produce further ion pairs
and the total ionization differs from the primary ionization (see factor
$\eta _1$ below).

The primary ionization rate
per unit volume per unit time interval
caused by \kr protons of differential number density $n_{\rm p}(T_{\rm p})$ is
then obtained by integrating the average ionization loss rate of \kr
 protons from Eq.(4.32) multiplied with $n_{\rm p}(T_{\rm p})$ over all
kinetic energies and dividing by 32~eV, yielding
$$f_{CR}=\int_{13.6 eV}^\infty dT_{\rm p}
\left|{dE\over{dt}}(T_{\rm p})\right|
{n_{\rm p}(T_{\rm p})\over 32 {\rm eV}}\eqno (4.41).$$
The total \kr ionization rate is then
$$\xi_{CR}=\eta _1\eta _2\eta _3f_{CR}/
n_{\rm H}\eqno (4.42),$$
where according to Takayanagi (1973)
$\eta _1=5/3$ accounts for secondary ionization from electrons
produced in the ionization by the primary \kr particle (Dalgarno and Griffing
1958), $\eta _2\simeq $1.17 accounts for a 10$\% $ helium abundance
by number density in the interstellar medium, and $\eta _3\simeq $1.43
accounts for the contribution of
heavy \kr \nuci and \kr electrons and positrons. Strictly, Eqs.(4.41)
and (4.42) apply only for the ionization of
neutral HI gas. However, it may also be used for estimating the ionization
rate in a completely ionized HII gas (Spitzer 1948, Ginzburg 1969).

The heating rate $\Gamma _{CR}$
of matter by cosmic rays is obtained by multiplying
the total ionization rate (4.42) with the matter density and the
energy $\Delta Q\simeq 20$ eV that is finally deposited as heat
from each ionization process (Spitzer and Scott 1969, Goldsmith and
Langer 1978). With Eq.(4.41) we then obtain
$$\eqalignno{&\Gamma _{CR}=\Delta Q\xi _{CR}n_{\rm H}&(4.43)\cr &
=(5/8)\eta _1\eta _2\eta _3
\int_{13.6 eV}^\infty dT_{\rm p}\left|{dE\over {dt}}(T_{\rm p})\right|
n_{\rm p}(T_{\rm p}) \cr}.$$
The evaluation of Eqs.(4.42) and (4.43) in the Galaxy requires the knowledge
of the interstellar \kr proton energy spectrum down
to energies of 13.6 eV, in a range
where it is not accessible to direct observations because of the
solar modulation.
Extrapolation of the known high-energy \kr \nuci spectrum
to lower energies adopting a straight power law injection
spectrum yields significant interstellar ionization and heating rates
up to $10^{-25}$~erg~cm$^{-3}$~s$^{-1}$ can be
attained
(see e.g. Lerche \& Schlickeiser 1982).
\titleb{Continuum radiation processes of relativistic \nuci }
Cosmic ray \nuci in distant part of the Universe may be detected
by their radiation products photons and neutrinos.  Low
energy neutrino astronomy
has just started to provide detectable fluxes with the eight neutrinos
measured at the explosion of SN 1987A in the Large Magellanic Cloud with
the Kamiokande neutrino detector.

At very high energies neutrino astronomy is
particulary interesting, because a) the solar and
atmospheric background are very weak (the latter because the atmosphere acts as
a thick
target for the high energy cosmic rays steepening the neutrino spectra from the
showers)
and b)
the detection probability increases with energy due to the
increasing neutrino--matter cross section (Stenger {\it et al.} 1992).

Electromagnetic radiation from \kr \nuci lies in the gamma ray band
and results from excitation of target nuclei by subrelativistic
\nuci (Section 4.4) and from neutral pion production by 1--30 GeV \nuci
(Section 4.2).
The pion decay radiation
spectrum  reveals a remarkable symmetry around
$(1/2)m_{\pi ^o}c^2\simeq 70$~MeV, and detection of this characteristic
spectrum is a strong argument in favor of a baryonic \kr origin of this
radiation. However, \kr \nuci also produce secondary charged pions
which decay into electrons and positrons, the bremsstrahlung of which
is also emitted in the MeV--GeV photon energy range and which may mask the
$\pi ^o-$bump at 70~MeV . So nondetection of the
$\pi ^o-$bump at 70~MeV would not naturally argue against a nucleonic
origin of these gamma rays, but would indicate the presence of many
electrons or positrons (for further
discussion of this point see Schlickeiser 1982 and Pohl \& Schlickeiser
1991).
Nevertheless, the Compton Gamma Ray Observatory (CGRO) has detected a clear
signature of the pion hump
in the spectrum of the diffuse gamma rays in the Galaxy.
Hence, as nonthermal synchrotron radiation
is indicative for the presence of \kr electrons, gamma rays from
$\pi ^o$--decay as well as from bremsstrahlung radiation of secondary
electrons and positrons from charged pion decay are indicative for
the presence of {\kr \nuci}.

In the more exotic environments of
AGN and their associated jets  extremely high photon densities
have the consequence that
interactions of cosmic ray nuclei with photons may, in fact, be the dominant
source of high energy radiation.
Similar to cosmic rays hitting the Earth's atmosphere the energetic particles
can initiate showers in photon atmospheres (Berezinsky {\it et al.} 1990,
Mannheim 1993).

\titlea{Proton energy losses and equilibrium spectra}
The compilation of energy losses enables us
to discuss the stationary spectrum
of \kr protons in various types of galactic environments.

The transport equation of cosmic rays
neglecting Fermi--acceleration and galactic wind effects is given by
$$\eqalign{&\nabla\cdot \left[K({\bf p}, {\bf r})\nabla N({\bf p}, {\bf
r})\right]
+{1\over p^2}{\bf d\over dp}\left[-p^2\dot p_{\rm loss}(p)N({\bf p}, {\bf r})
\right]\cr &+{N({\bf p}, {\bf r})\over t_{\rm c}({\bf p})}=
S({\bf p}, {\bf r})\cr}\eqno(5.1)$$
({\it e.g.}, Wang
and Schlickeiser 1987), where $N({\bf p}, {\bf r})$ denotes
the particle's phase--space distribution.
Energetic particles are assumed to
be injected by a separable source\fonote{{}For a relaxation
of this assumption see Appendix of Lerche \& Schlickeiser (1988).}
$$S({\bf p}, {\bf r})=q({\bf r})Q({\bf p})\eqno(5.2).$$
The diffusion
coefficient is also assumed to be separable in momentum $\bf p$
and space $\bf r$, {\it viz.}
$$K({\bf p}, {\bf r})=k({\bf r})\kappa({\bf p})\eqno(5.3).$$
Continuous momentum losses
are denoted as $\dot p_{\rm loss}$.
All but the pion production energy losses are treated
as continuous
losses.  Pion production by collisions with the interstellar matter
has a large inelasticity $K_{\rm p}=0.5$ and therefore effectively
removes particles from phase space in the case of steep power law spectra.
Therefore a catastrophic
loss time scale $t_{\rm c}=T_{\rm p}|dE_{\rm p}/dt|_\pi^{-1}$
(Eq. 4.11) is introduced in Eq. (5.1).

As shown in Wang and Schlickeiser (1987)
it is possible to decouple spatial and momentum transport,
respectively, by the transformation
$$N({\bf p}, {\bf r})=\int_0^\infty duF({\bf p}, u)T({\bf r},u)\eqno(5.4),$$
where $T({\bf r},u)$ with the normalization
$\int_0^\infty du T({\bf r},u)=1$
is commonly referred to as the age distribution
of cosmic rays at position ${\bf r}$.
The age distribution of cosmic rays is then given by
an infinite series of eigenfunctions
$$T({\bf r},u)=\sum_{l=0}^\infty \sum_{m=1}^\infty A_{lm}({\bf r})
\exp\left(-\omega_{lm}^2u\right)\eqno(5.5)$$
yielding the phase--space distribution
$$N({\bf p},{\bf r})=\sum_{l=0}^\infty\sum_{m=1}^\infty
A_{lm}({\bf r})N_{lm}({\bf p})\eqno(5.6),$$
where $N_{lm}({\bf p})$ obeys the differential equation
$$\eqalign{&{1\over \kappa(p)p^2}{d\over dp}
\left[p^2\dot p_{\rm loss}(p)N_{lm}(p)\right]\cr &+
\left[{1\over \kappa(p)t_{\rm c}(p)}+\omega_{lm}^2\right]N_{lm}(p)=
{Q(p)\over \kappa(p)}\cr}\eqno(5.7)$$
which has been referred to as the
leaky--box equation ({\it e.g.} Jones 1970).
Transforming to the particle number density
$n_{\rm p}(p)=4\pi p^2 N(p)$ gives
$$\eqalign{&-{\partial\over \partial p}\left(\sum_i|\dot p_i(p)|
n_{\rm p}(p)\right)\cr &+n_{\rm p}(p)\left[{1\over t_{\rm esc}(p)}+
{1\over t_{\rm c}(p)}\right]=
q_{\rm p}(p)\cr}\eqno(5.8),$$
where $q_{\rm p}(p)$ denotes the volume emissivity of sources.
The diffusive escape--time corresponding to
the lowest order eigenfunction is then given by
$$t_{\rm esc}(p)=
{1\over \omega_{01}^2\kappa(p)}\eqno(5.9).$$
Empirically, from measurements of cosmic ray Be$^{10}$--isotope
abundance in
the Galaxy,  Garcia--Munoz {\it et al.} (1977) obtain
$$t_{\rm esc}(p)=t_0\left(p/\rm GeVc^{-1}\right)^{\rm -w}
\eqno(5.10),$$
where $t_0\approx 2\cdot 10^7$~yrs
and
$\rm w\approx 0.5$.
A commonly used source of energetic cosmic ray protons in the
Galaxy is a power law
$$ q_{\rm p}(p)=q_\circ p^{\rm -s}\eqno(5.11)$$
with $\rm s=2.2$  as expected for Fermi--acceleration at shocks.

Imposing the boundary value
$n_{\rm p}(\infty)\rightarrow 0$
we obtain the stationary cosmic ray spectrum in terms of
kinetic energy $T_{\rm p}$ (as an implicit argument)
$$\eqalignno{&
n_{\rm p}(T_{\rm p})
=
|\dot E(T_{\rm p})|_{\rm tot,Z=1}^{-1}&(5.12)
\cr &\cdot\int_{T_{\rm p}}^\infty dT_{\rm p}'q_{\rm p}(T_{\rm p}')
\exp\left[-\int_{T_{\rm p}}^{T_{\rm p}'}{dT_{\rm p}''[t_{\rm esc}^{-1}+
t_{\rm c}^{-1}] (T_{\rm p}'')\over
|\dot E(T_{\rm p}'')|_{\rm tot,Z=1}}\right]\cr}
$$
including the effects of
ionization,
Coulomb scattering and
proton--matter hadronic interactions according to Eq.(4.40), as well as
pair production and photo--hadron production
with respect to nonthermal  and
thermal target radiation fields according to Eqs.(3.6), (3.6), (3.8) and (3.9).
\titleb{Examples}
The stationary proton distribution (5.12) has a particularly simple
interpretation.  Given a proton injection
with a straight power law in momentum, {\it i.e.}
$q_{\rm p}[p(T_{\rm p})]\propto p^{-s}$, (5.12) entails
the deviations from this power law
due to the various energy loss processes and diffusive escape.
We have $n_{\rm p}\propto t_{\rm loss}q_{\rm p}$ and because
of $t_{\rm loss}=[\sum_{\rm i} t_{\rm i}^{-1}]^{-1}$ the energy dependence
of the process with the shortest time scale in a given energy range determines
the final slope.  For example, $t_{\rm pp}\propto const.$ and hence
$n_{\rm p}\propto T_{\rm p}^{-s}$ when proton-proton collisions dominate.
On the other hand, for $T_{\rm p}\ll m_{\rm p}c^2$ the momentum vs. kinetic
energy relation $p=\sqrt{T_{\rm p}^2+2T_{\rm p}m_{\rm p}c^2}$ yields
$p\propto T_{\rm p}^{1/2}$, so that there is flattening in the non-relativistic
regime in addition to the flattening caused by the energy dependence
of the ionization losses $t_{\rm ion}\propto T_{\rm p}^{3/2}$ (above
$T_\circ$ Eq. 4.28), {\it viz.} $n_{\rm p}\propto T_{\rm p}^{-{1\over 2}
(s-3)}$.

Since the combined Coulomb and
ionization losses and the proton-proton hadronic
interaction losses are both proportional to the matter density $n_{\rm H}$,
the kinetic energy where both are equal depends only on natural constants
and is given approximately by $T^*=450$~MeV.  Therefore, a universal steepening
of the cosmic ray proton spectra from $n_{\rm p}\propto T_{\rm p}^{-
{1\over 2}(s-3)}$ to $n_{\rm p}\propto T_{\rm p}^{-s}$ is expected.
However, the escape time scale may well be shorter than the proton-proton
interaction time scale at $T^*$, so that the steepening is actually more
severe, {\it viz.} $n_{\rm p}\propto T_{\rm p}^{-(s+w)}$.
The cosmic rays lost through diffusive escape do not contribute to
the gamma ray emissivity.

Let $T^{**}$ denote the proton kinetic
energy where escape losses win over proton-proton
interaction losses.  Then, if $T^{**}>T^*$ holds, a galaxy would be
a very strong gamma ray emitter, since then the bulk of cosmic ray protons
would produce pions before being lost from the system.  This could be
the case in galaxies where the gas density is very large (Fig.1).
However,
for our Galaxy, conversely $T^{**}<T^*$ holds (Fig.1), so that as seen from
outside of the Galaxy a greater flux of cosmic ray protons
than  gamma rays is emitted.
\begfig8cm
\figure{2}{Same as Fig.~1, but for the case of the central region
of a radio quiet AGN.  Cosmic rays are likely to be produced there.
However, before they reach the tenuous interstellar matter where
diffusive transport dominates they would inevitably be damped by
the strong radiation field.
A photon luminosity of
$L=10^{44}$~erg/s and a thermal spectrum peaking
at $kT=30$~eV within a sphere of radius $R=10^{-3}$~pc was assumed.
Further assumptions are that the plasma in the central region
has a density of
$n_{\rm e}=3\cdot 10^5$~cm$^{-3}$ and is
heated to the
Compton temperature of $T_{\rm C}=10^7$~K .  The magnetic field energy density
is assumed to be
in equipartition with the radiation field.  Escape has been assumed to
be negligible compared to the energy loss processes.}
\endfig
A special situation occurs in
galaxies with active galactic nuclei (AGN).  They
have very luminous nuclei and show direct evidence for the presence of
relativistic particle populations.
The strong EUV radiation field
with a mean photon energy
$\langle \varepsilon_\gamma\rangle\approx 50$~eV characteristic for most AGN
is probably caused by
an accretion disk around a supermassive Black Hole.
Cosmic ray protons exceeding the threshold
energy for pion production (2.14) $\gamma_{\rm p,\pi}\simeq 10^6$
and pair production (2.15)
$\gamma_{\rm p,e^\pm}\simeq 10^4$, respectively,
cool mainly on the ambient photon target
when $n_\gamma\ge  n_{\rm H}/\alpha_{\rm f}$ and
$n_\gamma\ge  21 n_{\rm H}/\alpha_{\rm f}$, respectively,
where $\alpha_{\rm f}=1/137$.
The number density of the target photons is
$$\eqalign{&n_\gamma={L\over 4\pi R^2 c \langle\varepsilon\rangle}\cr &=3\cdot
10^{11}\left(R\over 10^{-3}
\thinspace{\rm pc}\right)^{-2}\left(L\over 10^{44}\thinspace{\rm erg/s}\right)\
{\rm cm}^{-3}\cr}\eqno(5.14),$$
whereas pressure equilibrium
with line emitting clouds
gives some $n_{\rm H}\le 3\cdot 10^5$~cm$^{-3}$ for the matter density in
the central regions
of AGN. Thus, within the central tenths of a parsec cosmic ray nuclei at
very high energies must cool
dominantly via photoproduction of secondaries (Fig.~2).
If the cosmic ray nuclei injection has a slope $s\le 2$, this cooling
channel would cause very powerful high energy electromagnetic cascades.
On the other hand, the cosmic ray spectrum in the outer regions of the
host galaxies harbouring the AGN would be devoid of particles at very
high energies.  Of course,  other cosmic ray production
sites such as supernova remnants, stellar winds and pulsars would
have to be considered also, but  presumably they are much weaker
than the AGN.

The photons in radio loud AGN typically have a nonthermal energy distribution
$n_\gamma(\varepsilon)\propto \varepsilon^{-2}$ corresponding to equal
power per logarithmic bandwidth.   Due to the inverse photon
distribution protons find more and more target photons for photoproduction
$n_\gamma\propto \int d\varepsilon n_\gamma
(\varepsilon)\propto \varepsilon^{-1}$  as their energy
increases above threshold, {\it i.e.} the energy loss increases as
$dE/dt\propto E^2$ and
the corresponding stationary proton distribution steepends to $s+1$ (Fig.~3).
\begfig8cm
\figure{3}{Same as Fig.~2, but for the case of a compact region in the
jet of a radio loud AGN with
rest frame luminosity $L=10^{44}$~erg/s and a nonthermal
continuum spectrum with spectral index $\alpha=1$.}
\endfig
\titleb{The cosmic gamma ray background}
Apart from the question whether the volume emissivity of any known source
population of
gamma rays contributing to the cosmic gamma ray background (CGB) is
large enough to explain
the observed omnidirectional flux ({\it e.g.} Fichtel 1983),
one may ask whether the observed spectrum can be reconciled
in a simple manner with a \kr origin in external galaxies.  Here it is
interesting to compare
the rest frame gamma ray spectra of individual galaxies for the cases sketched
above.  That
is (i)  spirals as our own Galaxy
with $t_{\rm pp}> t_{\rm esc}$  (ii)
gas rich galaxies with $t_{\rm pp}<t_{\rm esc}$ over a broad energy range
(cf. Ormes {\it et al.} 1987) and (iii) ultraluminous
galaxies with jets and active nuclei that could produce
{\krs} cooling {\it in
situ} by photoproduction of secondaries before diffusion spreads them
out in the interstellar medium of the host galaxy, such that
$t_{\rm p\gamma}<t_{\rm pp}<t_{\rm esc}$.

Assuming that the
\kr spectrum injected from the sources
into the interstellar medium has a slope of
$s\approx 2$ and that energy-dependent escape losses
steepen the spectrum by a power of $0.5$
type (i) spectra have a slope $s_\gamma\approx 2.5$ above $70$~MeV.
Below the slope is
the same due to bremsstrahlung from charged pion decay pairs
and primary \kr electrons.   Since for such a spectrum
the luminosity increases $\propto \varepsilon^{-1/2}$
but is limited, the slope must flatten towards lower photon energies.  Indeed,
the
CGB exhibits a bump in
$\nu S_\nu$
at roughly $3$~MeV.
It remains to be shown, whether bremsstrahlung can produce the bump for
moderate redshifts
$z=2-5$ in detail.  Alternatively, this can
only be identified with the neutral pion bump for extremely large redshifts
$z\approx 20$.
However, still their remains the problem that the present-epoch volume
emissivity
of normal spirals seems to be insufficient.  The fact that $t_{\rm pp}>t_{\rm
esc}$ means,
however, that even if $L_{\gamma}=0.5u_{\rm p}V/
t_{\rm pp}$ is small the corresponding luminosity of escaping \kr protons can
be large
$L_{\rm p}=2L_\gamma t_{\rm pp}/t_{\rm esc}$.  The escaping flux of \krs can be
an important
source of ionization of the intergalactic matter.

Type (ii) spectra have a flatter gamma ray
slope $s_\gamma\approx 2$ from $70$~MeV up to a few hundred MeV and then
steepen above.
The bremsstrahlung component in the MeV range should behave similar leading to
a more
pronounced bump at a few MeV than for case (i).  It is unclear whether there is
a numberous
source population for this type.  Due to obscuration it might be undiscovered
so far.
FIR observations seem to indicate that indeed a large number of faint sources
exists at
great redshifts with unknown high energy spectra.
Type (iii) is certainly an important contributor to the CGB.
Radio loud AGN are the most active population of extragalactic gamma ray
point sources as shown by the EGRET experiment on board CGRO.
Although some show spectra with $s_\gamma=2.5$ it is not clear whether
the sum of all spectra can account for the CGB spectrum
(Dermer \& Schlickeiser 1992, Stecker {\it et al.} 1993).
The emerging gamma ray spectrum from \kr
sources sitting
inside photon atmospheres rather lets one expect
an average spectral index of
$s_\gamma\approx 2$ below the energy where $\tau_{\gamma\gamma}=1$
due to cascade reprocessing.  At present it is unclear where
this energy lies on average.
Moreover, the bump at 3~MeV is not naturally explained.
However, the bump power is surprisingly similar to the 40~keV
bump of the CXB suggesting a common physical origin, such as the radio quiet
AGN.
The characteristic energy of the CXB is 40~keV corresponding to 120~keV in the
rest frame
of the AGN at $z=2$.  This can well be accounted for by Compton-reflection of
hard X-rays on cold
matter.  What is the origin of the corresponding characteristic
energy 10~MeV for the CGB
in this case?

\titlea{Summary}
We have presented a collection of formulae for the
energy loss of \kr nuclei over a vast range of nucleus kinetic energies.
The processes considered are photo--pair and -hadron production,
photodisintegration,
proton-matter pion production, fragmentation, Coulomb scattering and
ionization.
Results are applied to the problems of ISM heating and gamma ray spectra from
galaxies.

At low energies ionization losses depopulate the stationary \kr energy
distribution below
the universal kinetic energy
$T^*\approx 450$~MeV.
The absorbed energy ultimately heats the ISM at a considerable rate,
although other heat sources such as young stars may well dominate.  However,
some galaxies
have an enhanced nonthermal activity (as compared to the stellar emission), so
that \kr heating very likely
becomes the dominant heat source.

Above $T^*$ pion production is the more important process which does
not change the spectral index of the injected \kr source spectrum and leads to
gamma ray emission
by neutral pion decay and bremsstrahlung from pairs resulting from charged pion
decay.
Diffusive escape of cosmic rays from the galactic containment volume steepens
the slope of the \kr
spectrum inside the galaxy, so that the slope of the gamma ray spectrum attains
values $\approx 2.5$
for injection with slope $\approx 2$.  Diffusion coefficients different from
that for
\kr transport in our Galaxy would lead to a different steepening of the
spectrum.  Apart from
that
an enhanced gas density $n_{\rm H}>1$~cm$^{-3}$
can flatten the gamma ray spectrum above $70$~MeV.  Without invoking redshifts
$z_{\rm max}\approx 24$
neutral pion decay gamma rays can not explain the observed bump
of the CGB at $3$~MeV.  To explain the CGB with \kr induced gamma rays one must
describe the \kr electrons and their bremsstrahlung spectrum as well.

A special situation can occur in galaxies with active nuclei.   The energetic
photons in their
strong nonstellar radiation fields outnumber the atoms of the interstellar
matter
by a large margin.   Photoproduction of secondaries then becomes the dominant
energy loss
process at high energies leading to strong gamma ray emission with complex
spectra.
It is possible that the CGB above $100$~MeV
can be explained by the gamma rays from radio loud
AGN ({\it e.g.}, Dermer \& Schlickeiser 1992 and Stecker {\it et al.} 1993
argue for a
contribution of at least 20\%).  However, due to the relativistic beaming
of the jet emission an assessment of their total contribution is difficult
to achieve.
The fact that the 3~MeV bump power is similar to the CXB bump power at 40~keV
makes
radio quiet AGN
as a common physical origin plausible, but it
does not lead to a {\it simple} explanation of the characteristic
CGB energy of roughly 10~MeV ($z_{\rm max}\simeq 2$) without invoking rather
specific
model scenarios of AGN.

\acknow{We thank the referee for his very pertinent comments.
R.S. thanks for support of his GRO
guest investigator program by DARA (50 OR 9301 1).}

\begref{References}

\ref  Begelman, M.C., Rudak, B., Sikora, M., 1990,  {\it Astrophys. J.},  {\bf
362}, 38

\ref Berezinsky, V.S., Grigoreva, S.I., 1988, {\it Astron. Astrophys.}, {\bf
199}, 1

\ref Berezinsky, V.S., Bulanov, S.V., Dogiel, V.A., Ginzburg, V.L. (ed.),
Ptuskin,
1990, in {\it Astrophysics of Cosmic Rays}, North-Holland, Amsterdam

\ref Bethe, H.A., Heitler, W., 1934, {\it Proc. Roy. Soc.} (London), {\bf
A146}, 83

\ref Biermann, P.L., Strittmatter, P.A., 1987, {\it Astrophys. J.}, {\bf 322},
643

\ref Blumenthal, G.R., 1970, {\it Phys. Rev. D} {\bf 1}, 1596

\ref Brown,M.D.,  Moak,C.D., 1972, {\it Phys. Rev. B}, {\bf 6}, 90

\ref Butler, S.T., Buckingham, M.J.,  1962, {\it Phys. Rev.}, {\bf 126}, 1

\ref Cheng, C.-C., 1972, {\it Space Sci. Rev.},  {\bf 13}, 3

\ref Chodorowski, M.J., Zdziarski, A.A., Sikora, M., 1992, {\it Astrophys. J.},
{\bf 400}, 181

\ref Dalgarno, A., Griffing, G.W., 1958, {\it Proc. R. Soc. London A}, {\bf
248}, 415

\ref Dermer, C.D., 1986, {\it Astron. Astrophys.}, {\bf 157}, 223

\ref Dermer, C.D., Schlickeiser, R., 1992, {\it Science}, {\bf 257}, 1642

\ref Evans, R.D., 1955, {\it The Atomic Nucleus}, McGraw-Hill, New York

\ref Fermi, E., 1950, {\it Prog. Theo. Phys.}, {\bf  5}, 570


\ref Fichtel, C.E., 1983, {\it Adv. Space Res.}, Vol.3, No.4, 5

\ref Garcia-Munoz, M., Mason,G.M., Simpson, J.A., 1977, {\it Astrophys. J.},
{\bf 217}, 859

\ref Genzel, H., Joos, P., Pfeil, W., 1973, in: {\it Landolt--B\"ornstein},
Vol. 8,
Ed. Schopper, H., Springer, Berlin, p. 16


\ref Ginzburg, V.L., 1969, {\it Elementary Processes for Cosmic Ray
Astrophysics},
Gordon and Breach, New York

\ref Ginzburg, V.L., Syrovatskii, S.I., 1964, {\it The Origin of Cosmic Rays},
Pergamon, New York

\ref Goldsmith, P.F., Langer, W.D., 1978, {\it Astrophys. J.}{\bf  222}, 881

\ref Gould, R.J., Burbidge, G.R., 1967, in: {\it Hand\-buch der Physik},
Vol.46/2,
Ed. Sitte, K., Springer, Berlin, p. 265

\ref Gould, R.J., 1972, {\it Physica},  {\bf 62}, 555

\ref Greisen, K., 1966, {\it Phys. Rev. Lett.}, {\bf 16}, 748

\ref Hagedorn, R., 1973, {\it Relativistic Kinematics}, Benjamin, Reading

\ref Hayakawa, S., 1969, {\it Cosmic Ray Physics}, John Wiley, New York

\ref Heitler, W., 1954, {\it The Quantum Theory of Radiation}, Oxford
University Press,
Oxford

\ref Jones, F.C., 1970, {\it Phys. Rev. D}, {\bf 2}, 2787

\ref Lerche, I., Schlickeiser, R., 1982, {\it Mon. Not. R. astr. Soc.},
{\bf 201}, 1041

\ref Lerche, I., Schlickeiser, R., 1988, {\it Astrophys. and Sp. Sc.},  {\bf
145}, 319

\ref Letaw, J.R., Silberberg, R., Tsao, C.H., 1983,
{\it Astrophys. J. Suppl.}, {\bf  51}, 271

\ref Mannheim, K., Biermann, P., 1989, {\it Astron. Astrophys.}, {\bf 221}, 211



\ref Mannheim, K., 1993, {\it Astron. Astrophys.}, {\bf 269}, 67

\ref Marscher, A.P., Vestrand, W.T., Scott, J.S., 1980, {\it Astrophys. J.},
{\bf 241}, 1166


\ref Meneguzzi, M., Reeves, H., 1975, {\it Astron. Astrophys.}, {\bf 40}, 91

\ref Murphy, R.J., Dermer, C.D., Ramaty, R., 1987,
{\it Astrophys. J. Suppl.}, {\bf 63}, 721

\ref Northcliffe, L.C., 1963, {\it Ann. Rev. Nucl. Sci.}, {\bf  13}, 67

\ref Ormes,J.F., \"Ozel, M.E., Morris, D.J., 1987, {\it Astrophys. J.}, {\bf
334}, 722

\ref Perola, G.C., Scarsi, L., Sironi, G., 1967, {\it Nuovo Cimento}, {\bf
52B}, 455

\ref Pohl, M., Schlickeiser, R., 1991, {\it Astron. Astrophys.}, {\bf 252}, 565

\ref Puget, J.L., Stecker, F.W., Bredekamp, J.H., 1976, {\it Astrophys. J.},
{\bf 205}, 638

\ref Ramaty, R., Lingenfelter, R.E., 1966, {\it J. Geo. Res.}, {\bf 71}, 3687

\ref Ramaty, R., 1974, in: {\it High Enery Particles and Quanta in
Astrophysics},
eds. F.B. McDonald and C.E. Fichtel, MIT Press, Cambridge, p. 122

\ref Ramaty,R., Kozlovsky, B., Lingenfelter, R.E., 1979, {\bf ApJS} 40, 487


\ref Scanlon, J.H., Milford, S.N., 1965, {\it Astrophys. J.}, {\bf 141}, 718

\ref Schlickeiser, R., 1982, {\it Astron. Astrophys.}, {\bf 106}, L5

\ref Sikora, M., Kirk, J.G., Begelman, M.C., Schneider, P., 1987,
{\it Astrophys. J.}, {\bf 320}, L81

\ref Silberberg, R., Tsao, C.H., 1973, {\it Astrophys. J. Suppl.}, {\bf 25},
315

\ref Silberberg, R., Tsao, C.H., 1990, {\it Phys. Rep.}, {\bf  191}, 351

\ref Sivukhin, D.V., 1965, in: {\it Rev. Plasma Physics}, ed. M. Leontovich,
Consultans Bureau, New York

\ref Sorrell, W.H., 1987, {\it Astrophys. J.}, {\bf 323}, 647

\ref Spitzer, L.Jr., 1948, {\it Astrophys. J.}, {\bf 107}, 6

\ref Spitzer, L.Jr., Scott, E.H., 1969, {\it Astrophys. J.}, {\bf 158}, 161

\ref Stecker, F.W., 1968, {\it Phys. Rev. Lett.}, {\bf 21}, 1016

\ref Stecker, F.W., 1969, {\it Phys. Rev.}, {\bf 180}, 1264

\ref Stecker, F.W., 1973, {\it Astrophys. J.}, {\bf 185}, 499

\ref Stecker, F.W., Salamon, M.H., Malkan, M.A.,
1993, {\it Astrophys. J.}, {\bf 410}, L71

\ref Stenger, V.J., Learned, J.G., Padvasa, S., Tata, X. (eds.),
1992, {\it High Energy Neutrino Astrophysics}, Proc. of the Workshop
held in Honolulu, Hawaii, 23-26 March 1992, World Scientific, Singapore

\ref Sternheimer, R.M., 1952, {\it Phys. Rev.}, {\bf  88}, 851

\ref Takayanagi, K., 1973, {\it PASJ}, {\bf  25}, 327

\ref Wang, Y.-M., Schlickeiser, R., 1987, {\it Astrophys. J.}, {\bf 313}, 200

\ref Whaling, W., 1958, in: {\it Hand\-buch der Physik}, ed. S. Fl\"ugge,
Springer,
Berlin, Vol. 34, p. 193

\ref Zatsepin, G.T., Kuzmin, V.A., 1962, {\it Soviet Phys. JETP}, {\bf  14},
1294


\endref
\end